\documentclass[lettersize,journal]{IEEEtran}
\usepackage{iftex}
\ifPDFTeX\else
  \PackageError{bare_jrnl_new_sample4}{Use pdfLaTeX, not XeLaTeX or LuaLaTeX}{Select the LaTeX Workshop recipe: pdflatex-shellescape -> bibtex -> pdflatex x2.}
\fi
\usepackage{amsmath,amsfonts}
\usepackage{algorithmic}
\usepackage{algorithm}
\usepackage{array}
\usepackage[caption=false,font=normalsize,labelfont=sf,textfont=sf]{subfig}
\usepackage{textcomp}
\usepackage{stfloats}
\usepackage{url}
\usepackage{verbatim}
\usepackage{graphicx}
\usepackage{cite}
\DeclareUnicodeCharacter{03BB}{$\lambda$}
\DeclareUnicodeCharacter{00D7}{$\times$}
\def\BibTeX{{\rm B\kern-.05em{\sc i\kern-.025em b}\kern-.08em
    T\kern-.1667em\lower.7ex\hbox{E}\kern-.125emX}}
\usepackage{balance}
\begin{document}

\title{Slim-Mamba: An Efficient and Reconfigurable Mamba-Based Accelerator for Massive MIMO Indoor Localization}

\author{Shengjie Chen$^{\dagger}$, Chenxi Zhang$^{\dagger}$, Sijia Cheng, Ove Edfors, Liang Liu, and Ilayda Yaman 
\thanks{This work is partially funded by the Swedish Research Council, and ELLIIT (Excellence Center at Linköping-Lund in Information Technology), and partially supported by the Competence Center NextG2Com funded by the VINNOVA program for Advanced Digitalisation with grant number 2023-00541. \textit{Corresponding author: Ilayda Yaman.}

The authors are with Department of Electrical and Information Technology, Lund University, Sweden (email:{\tt\small name.surname@eit.lth.se}).}
 \thanks{$^{\dagger}$Shengjie Chen and $^{\dagger}$Chenxi Zhang contributed equally to this work.}
 
 \thanks{The complete implementation open source, including the algorithm, fixed-point software reference, and FPGA accelerator code at GitHub\cite{slim_mamba}.}
}



\maketitle

\begin{abstract}
Using real-world massive MIMO measurements from the LuViRA dataset~\cite{LuViRA2024}, we present a system-level solution combining Mamba-based model development and hardware acceleration for indoor localization. 
Three Mamba-based models are explored: the original Mamba architecture, CMamba, a channel-correlation-enhanced state-space model designed to exploit correlations across input channels, and our proposed Slim-Mamba. Hardware implementation and bottleneck analysis are then used to guide further model simplification and optimization. The resulting Slim-Mamba  simplifies the original Mamba selective state-space block into a lightweight gated recurrent update, preserving temporal state modelling and output gating while reducing the complexity of the selective-scan computation.
The deployed model contains 816k trainable parameters and has a model size
of 3.40\,MB under FP32 precision while maintaining high localization accuracy. The corresponding accelerator provides a reconfigurable architecture with a highly parallel MAC array optimized for matrix-vector operations and fine-grained pipelining, improving resource utilization and reducing inference latency. The hardware-aware fixed-point model achieves a mean localization error of 0.14\,m, closely matching the floating-point baseline. Implemented on a Zynq UltraScale+ ZCU102 FPGA, the four-block backbone has an estimated serial latency of 212.04\,\(\mu\)s per token at 100\,MHz.
\end{abstract}


\section{Introduction}
\IEEEPARstart{I}{ndoor} localization is becoming an important function in future wireless systems,
where communication, sensing, and positioning are expected to be more tightly
integrated. In the context of 6G, localization has been identified as both a key
service and an enabling technology for location-aware applications such as robotic
navigation, emergency healthcare, smart environments, and intelligent
transportation~\cite{Trevlakis2023}. These applications impose diverse accuracy requirements, ranging from meter-level positioning for general indoor tracking to sub-meter~\cite{Xiang2019Submeter} and
decimeter-level~\cite{Shi2023RTPhase} accuracy for emerging applications requiring more precise
spatial awareness. However, accurate indoor positioning
remains challenging because indoor radio propagation is affected by multipath
effects, noise, and the surrounding radio environment
~\cite{Zafari2019Indoor}.

Massive multiple-input multiple-output (MIMO) systems provide a promising measurement platform for radio-based indoor localization, offering rich spatial information together with delay-domain information obtained from channel impulse responses. However, the high dimensionality of massive MIMO channel features leads to substantial computational and memory requirements, making real-time neural network inference challenging under limited computational and memory resources.

Previous works have explored several machine learning (ML) methods for massive MIMO localization. 
Early studies employed fully connected neural networks (FCNNs) using spatial covariance matrices and truncated channel impulse responses (CIRs) as fingerprints, achieving centimetre-level positioning accuracy~\cite{TianICC2023}. However, the large fully connected layers result in model sizes that exceed the on-chip memory capacity of inference accelerators, and computational complexity remains a major challenge for training and hardware deployment~\cite{Tian2024}.
CNN-based approaches exploit the structured nature of channel state information (CSI) fingerprints and extract local spatial features through convolutional layers. An ASIP-based implementation combining a programmable vector processor with a dedicated CNN accelerator was presented in~\cite{AttariISCAS2021}, with a later design further extending the hardware platform~\cite{AttariASIP2025}.
More recently, Transformer-based models, built around self-attention, have been investigated in the field of localization. Self-attention can capture correlations between beam-domain 
channel responses, where pairwise similarity patterns provide valuable spatial information for 
user localization~\cite{Tian2024Attention}. However, the quadratic complexity of self-attention, together with its associated memory traffic remains a challenge for efficient hardware implementation~\cite{YamanTransformer2026}.

From the examples above, efficient deployment which reduces model complexity, computation, and 
memory movement while preserving accuracy still remains a critical challenge for
learning-based Massive MIMO localization. Mamba~\cite{Mamba2023}, based on a state-space model (SSM) that  replaces self-attention and conventional MLP blocks with selective state-space processing and gating, provides a
promising direction for this purpose. Mamba performs recurrent state updates with
input-dependent parameters while maintaining linear complexity with respect to
sequence length. This recurrent architecture enables it to explicitly explore the temporal information, which is attractive for indoor
localization where consecutive channel observations may provide useful temporal context for estimating the current position. Recent studies in other learning-based applications, such as anomaly detection, have also started to explore Mamba-based architectures alongside established CNN- and Transformer-based approaches~\cite{MambaAD2024}. However, the use of Mamba for radio-based localization remains relatively unexplored and comparisons of CNN-, Transformer-, and Mamba-based architectures for localization are still lacking. 

For hardware optimization, existing Mamba accelerators have explored computation reordering, quantization, pipelined execution dataflow, and hybrid dataflows~\cite{LightMamba2025,FastMamba2025,HCSA2025,MARCA2024}. These works suggest that the efficiency of Mamba acceleration depends on reusable hardware structures that maximize resource sharing, efficient pipelined dataflows that improve processing throughput and reduce inter-stage stalls, and quantization strategies that balance computational efficiency with inference accuracy. 

The Lund University Vision, Radio and Audio (LuViRA) dataset used in this paper provides synchronized vision, radio, audio, and motion-capture measurements for indoor localization research~\cite{LuViRA2024,LuViRAValidation2024}. For the radio modality, channel responses are measured using the Lund University massive MIMO (LuMaMi) testbed~\cite{LuMaMi2017}.
To tailor Mamba to radio-based indoor localization task and efficiently implement on hardware, we develop Slim-Mamba, an integrated Mamba-based localization model and FPGA accelerator. The Slim-Mamba model is designed and refined with hardware implementation constraints in mind, and its performance is compared with existing FCNN-, CNN-, and Transformer-based approaches. 
The proposed FPGA accelerator reuses computational resources across different projection layers, supports configurable execution at the Slim-Mamba block level, applies a unified requantization module, and uses fine-grained FIFO-based pipelining to reduce waiting time between projection, nonlinear activation, state update, and output gating.

The main contributions of this paper are summarized as follows:

\begin{itemize}
    \item Slim-Mamba, a lightweight Mamba-based model for massive MIMO indoor localization, is proposed. The model simplifies the selective state-space structure of the original Mamba formulation into a hardware-friendly gated recurrent update, reducing model complexity while achieving a mean localization error of 14\,cm.

    \item A reconfigurable FPGA accelerator is developed for Slim-Mamba. The architecture reuses shared MAC Array across global
    projection layers, supports configurable execution across Slim-Mamba blocks, employs a unified quantization module, and introduces fine-grained FIFO-based pipelining between local operations to reduce pipeline stalls and inference latency while improving hardware utilization.

    \item Slim-Mamba is evaluated using real-world indoor massive MIMO measurements from the LuViRA dataset and compared with existing FCNN-, CNN-, and Transformer-based localization methods. At 100\,MHz, the FPGA backbone requires an estimated 21,204 cycles per token across four blocks under serial execution.
    
\end{itemize}

\section{System Model and Measurement Campaign}
\label{sec:system_model}

\subsection{Measurement Scenario}

\begin{figure}[!t]
  \centering
  \includegraphics[width=0.98\linewidth]{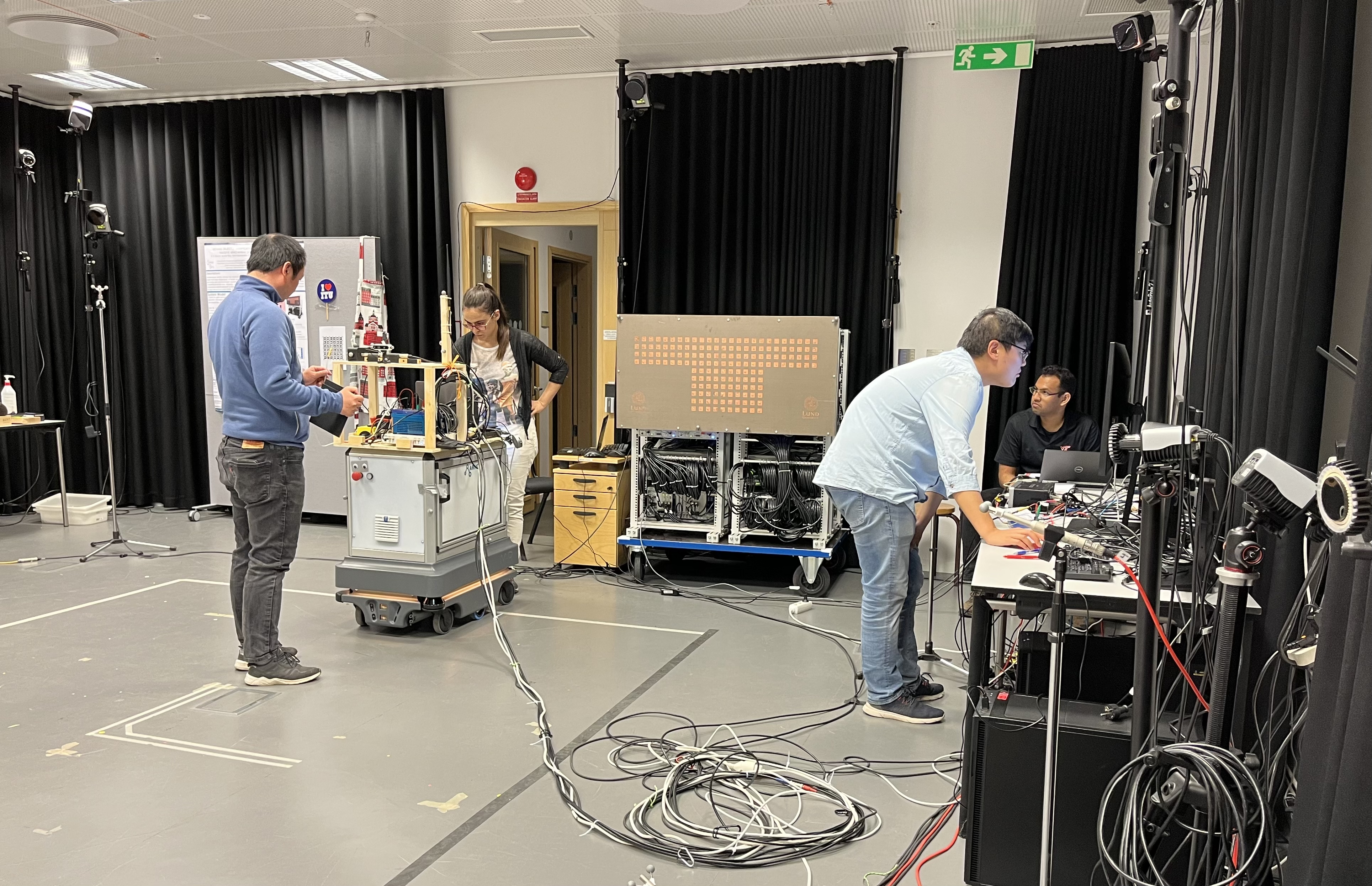}
  \caption{LuViRA measurement setup.}
  \label{fig:luvira_measurement_setup}
\end{figure}

We consider a single-input multiple-output (SIMO) uplink using orthogonal
frequency-division multiplexing (OFDM), from a single-antenna user equipment
(UE) mounted on a mobile robot to the 100-antenna LuMaMi receiver, as shown in
Fig.~\ref{fig:luvira_measurement_setup}. The LuViRA measurements were
collected at 3.7\,GHz over a 20\,MHz bandwidth, with 100 channel snapshots
per second. In the "grid" trajectories, the robot traversed approximately straight
paths at 0.1\,m/s, while a motion-capture system recorded the ground-truth
position~\cite{LuViRA2024,LuMaMi2017}.

\begin{figure}[!t]
  \centering
  \includegraphics[width=0.98\linewidth]{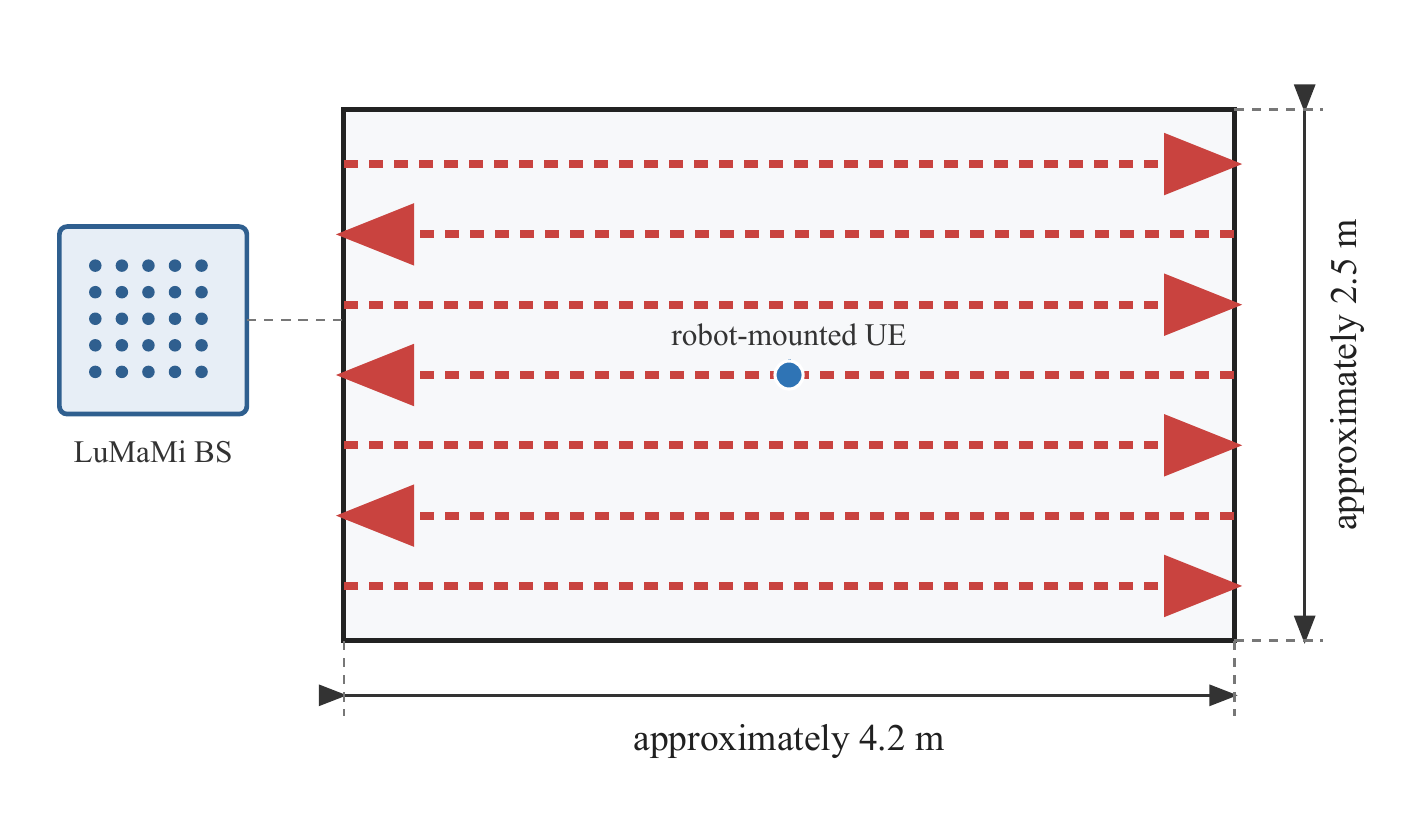}
  \caption{Grid-trajectory schematic, redrawn from~\cite{LuViRA2024,Tian2024}.}
  \label{fig:luvira_grid_scenario}
\end{figure}

Figure~\ref{fig:luvira_grid_scenario} illustrates the measurement geometry of the grid trajectories evaluated in this work. The robot-mounted UE moves along a dense raster of approximately parallel straight paths within the measurement area, with alternating traversal directions and about 0.05\,m spacing between adjacent paths. This geometry provides smooth local motion and dense spatial coverage, both of which can favor short-window sequence models.
Even-indexed grid trajectories form the held-out test set, while odd-indexed trajectories form the training pool. For Slim-Mamba, two odd-indexed trajectories are randomly selected from this pool for validation.

\subsection{Channel Features and Localization Metric}

We follow the truncated-CIR fingerprint construction in~\cite{Tian2024},
also used by the radio-localization baseline in~\cite{LuViRAValidation2024}.
Let $H_i[f,a]$ denote the measured frequency-domain channel at snapshot
$i$, frequency bin $f$, and receive antenna $a$. An inverse discrete
Fourier transform along the frequency axis produces the CIR; its first
$L=10$ delay taps are retained, as in~\cite{Tian2024}:
\begin{equation}
c_i[\ell,a]=\frac{1}{F}\sum_{f=0}^{F-1}H_i[f,a]
e^{j2\pi f\ell/F},\quad 0\leq\ell<L,
\label{eq:cir_features}
\end{equation}
where $F$ is the number of recorded frequency bins. For each trajectory
containing $T$ snapshots, the CIR is normalized separately for each antenna:
\begin{equation}
\tilde c_i[\ell,a]=\frac{c_i[\ell,a]}{\sqrt{\frac{1}{TL}
\sum_{j=1}^{T}\sum_{k=0}^{L-1}|c_j[k,a]|^2}+\epsilon},
\quad \epsilon=10^{-8}.
\label{eq:cir_normalization}
\end{equation}
The real and imaginary components of the normalized CIR
provide $2LM=2000$ features for $M=100$ antennas. The additional power
feature for each antenna is
\begin{equation}
q_i[a]=\log\!\left(1+\frac{1}{L}\sum_{\ell=0}^{L-1}
|c_i[\ell,a]|^2\right).
\label{eq:log_power_features}
\end{equation}
The log-power features retain the per-antenna power level removed by CIR
normalization, while the logarithm compresses its dynamic range. Concatenating
the CIR and log-power features gives $\mathbf{z}_i\in\mathbb{R}^{2100}$.
Each feature is standardized as
\begin{equation}
x_{i,d}=\frac{z_{i,d}-\mu_d}{\max(\sigma_d,\sigma_{\min})},
\quad d=1,\ldots,2100,
\label{eq:feature_standardization}
\end{equation}
where $\mu_d$ and $\sigma_d$ are training-set statistics and $\sigma_{\min}$
is the standard-deviation floor. These statistics are fixed during evaluation.
Both CMamba and Slim-Mamba use the resulting features.

For each position estimate, the model processes a temporal window of $K$
consecutive massive MIMO channel observations. The deployed configuration uses
\(K=2\). The input window
\(\mathbf{X}_t=[\mathbf{x}_{t-K},\ldots,\mathbf{x}_{t-1}]\)
is used to estimate the two-dimensional position $\mathbf{p}_t$ at the
following snapshot.

For an estimated position $\hat{\mathbf{p}}_i=(\hat p_{x,i},\hat p_{y,i})$
and its ground truth $\mathbf{p}_i=(p_{x,i},p_{y,i})$, the localization error is
\begin{equation}
e_i=\|\hat{\mathbf{p}}_i-\mathbf{p}_i\|_2
=\sqrt{(\hat p_{x,i}-p_{x,i})^2+(\hat p_{y,i}-p_{y,i})^2}.
\label{eq:localization_error}
\end{equation}
The mean localization error (ME) over $N$ evaluated positions is
\begin{equation}
\mathrm{ME}=\frac{1}{N}\sum_{i=1}^{N}e_i.
\label{eq:localization_me}
\end{equation}

\section{Mamba-Based Model Exploration for Localization}
\label{sec:algorithm}

\subsection{State-Space Models and Original Mamba}

For one scalar component $u(\tau)$ of the learned feature representation,
a continuous-time SSM maintains a latent state
$\mathbf{h}\in\mathbb{R}^{n_s}$:
\begin{equation}
\frac{d\mathbf{h}(\tau)}{d\tau}=A\mathbf{h}(\tau)+Bu(\tau),
\qquad r(\tau)=C\mathbf{h}(\tau)+Du(\tau).
\label{eq:continuous_ssm}
\end{equation}
Here $\tau$ denotes continuous time and $r(\tau)$ is the scalar state readout.
The input $u$ is a projected model feature derived from $\mathbf{x}_i$, whereas
$c_i[\ell,a]$ denotes a measured delay-domain channel coefficient.
The parameters $A$, $B$, $C$, and $D$ specify state evolution, input injection,
state readout, and the direct input path, respectively. Discretization with
step size $\Delta$ gives
\begin{equation}
\mathbf{h}_t=\bar A\mathbf{h}_{t-1}+\bar B u_t,
\qquad r_t=C\mathbf{h}_t+Du_t.
\label{eq:discrete_ssm}
\end{equation}
Mamba makes $B_t$, $C_t$, and $\Delta_t$ input-dependent while retaining
learned, input-independent $A$ and $D$~\cite{Mamba2023}. Its selective scan
evaluates the resulting time-varying recurrence and readout. Each feature
channel carries an $n_s$-dimensional state.

The original Mamba block in~\cite{Mamba2023} combines a local convolutional
front end with a selective state-space recurrence. After RMSNorm and input
projection, the input is split into a state-update path and a gate path. The
state-update path generates input-dependent parameters, performs
continuous-to-discrete conversion, and applies the selective scan, while the
gate path modulates the state output before the final output projection and
residual addition. This design supports general sequence modeling, but its
expanded recurrent state, input-dependent parameter generation, and
discretization increase state storage and operator diversity in a fixed-point
FPGA implementation.

We also adopted the CMamba architecture in~\cite{zeng2024cmamba} and retrained it for LuViRA with
task-specific input and output dimensions. Unlike the original Mamba block,
CMamba adds channel attention and training-time channel mixup to patch-based
Mamba processing, modeling interactions among projected feature channels.

\subsection{Slim-Mamba for Localization}

Slim-Mamba is designed to retain localization accuracy while simplifying
the recurrent datapath for FPGA deployment. It replaces the expanded
selective SSM with a channel-wise gated recurrence, reducing state storage
and the operators required for parameter generation and state readout.

\begin{figure}[!t]
  \centering
    \includegraphics[width=0.95\linewidth]{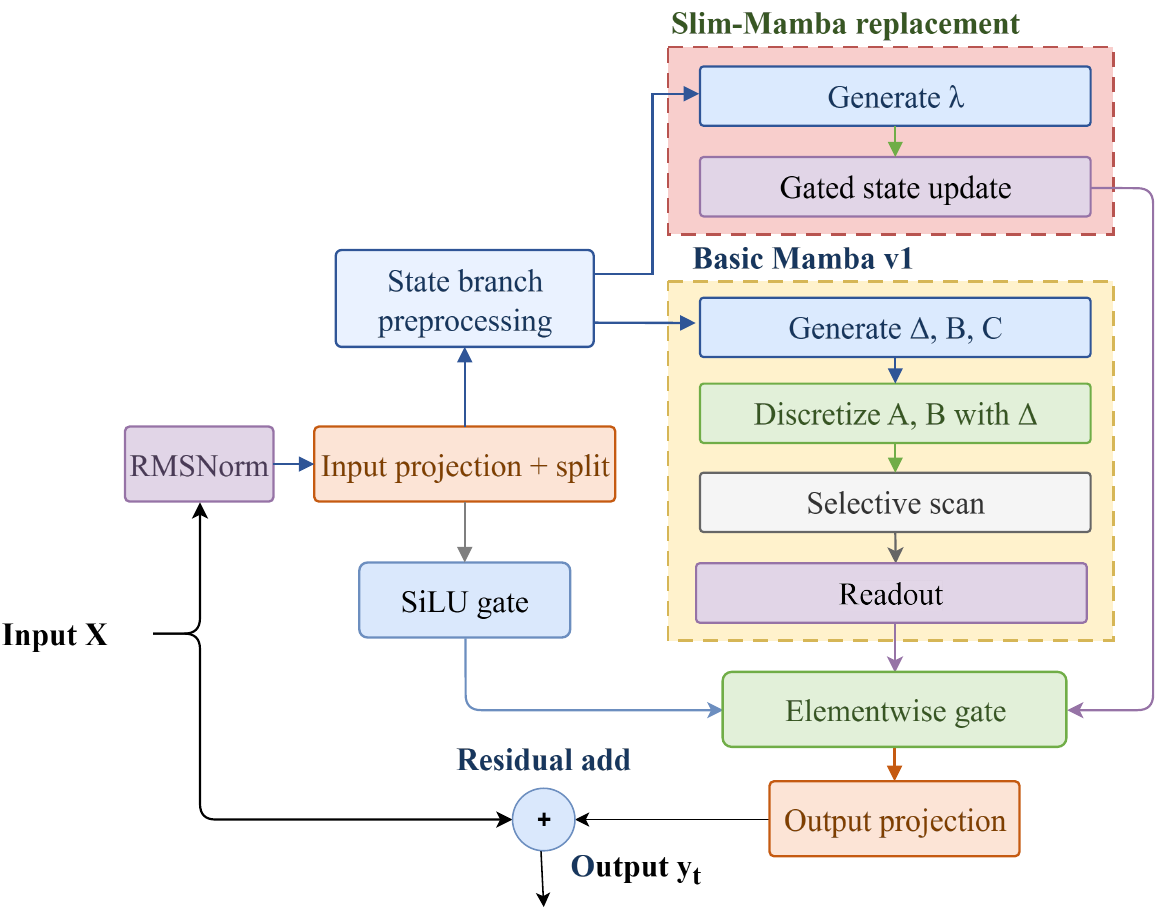}
    \caption{Structural comparison between the original Mamba block and the proposed
    Slim-Mamba block.}
    \label{fig:slim_mamba_replacement}
\end{figure}

As shown in Fig.~\ref{fig:slim_mamba_replacement}, Slim-Mamba retains the
overall block structure of original Mamba,
including RMSNorm, input projection and split, element-wise output gating,
output projection, and residual addition. The state-update path replaces
SSM parameter generation, discretization, and readout with a direct
discrete-time gated update. Let $\mathbf{b}_t$ denote an embedded block-input
token, with $t$ indexing tokens. After
\(\mathbf{b}_t\) passes through RMSNorm and the input projection-and-split
stage, the state-update path applies SiLU to produce
the feature \(\tilde{\mathbf{x}}^{(s)}_t\). The software model supports an
optional depthwise convolution before SiLU; it is disabled in the exported
deployment configuration. The update-coefficient projection applies weights
$W_{\lambda}$ and bias $b_{\lambda}$ to $\tilde{\mathbf{x}}^{(s)}_t$,
followed by a sigmoid to obtain the state-retention coefficient $\lambda_t$:
\begin{equation}
\lambda_t = \sigma \!\left(W_{\lambda} * \tilde{\mathbf{x}}^{(s)}_t + b_{\lambda}\right),
\label{eq:slim_mamba_lambda}
\end{equation}
where \(\sigma(\cdot)\) denotes the sigmoid function and $*$ denotes a
pointwise ($1\times1$) convolution. Instead of discretizing a
continuous-time state-space model, Slim-Mamba updates the recurrent state
directly as
\begin{equation}
\mathbf{h}_t = \lambda_t \odot \mathbf{h}_{t-1}
 + (1-\lambda_t)\odot \tilde{\mathbf{x}}^{(s)}_t.
\label{eq:slim_mamba_state_update_alg}
\end{equation}
This update has the complementary gating form derived for the scalar-state
selective SSM in~\cite[Theorem~1]{Mamba2023}. For a state dimension of one,
$A=-1$, $B=1$, and an input-dependent softplus step size, zero-order-hold
discretization yields complementary sigmoid coefficients for the previous
state and current input. In particular, let $a_t$ denote one component of
the update-coefficient projection output and set
$\Delta_t=\mathrm{softplus}(-a_t)$. Then
\begin{equation}
\bar A_t=e^{-\Delta_t}=\sigma(a_t),\qquad
\bar B_t=1-e^{-\Delta_t}.
\label{eq:scalar_ssm_gate_relation}
\end{equation}
Applying this scalar recurrence channel-wise with $\lambda_t=\sigma(a_t)$
yields Eq.~\eqref{eq:slim_mamba_state_update_alg}, with $\lambda_t$ controlling
state retention. Thus, the Slim-Mamba state update retains the gating
interpretation of this selective-SSM special case, using one state per
feature channel and direct state readout instead of the expanded state and
input-dependent readout of the general selective SSM. Computing $\lambda_t$
directly with sigmoid preserves this recurrence while avoiding explicit
softplus and exponential discretization operators.

The updated state is then modulated by the gate path
\begin{equation}
\mathbf{g}_t = \mathrm{SiLU}(\mathbf{x}^{(g)}_t),
\label{eq:slim_mamba_gate}
\end{equation}
and mapped to the block output through
\begin{equation}
\mathbf{y}_t = W_{\mathrm{out}}(\mathbf{h}_t \odot \mathbf{g}_t).
\label{eq:slim_mamba_output}
\end{equation}

Compared with the original Mamba block, the replacement in
Fig.~\ref{fig:slim_mamba_replacement} removes three major sources of hardware
complexity: time-varying generation of multiple SSM
parameters, the discretization step from \(\Delta_t\) to
\((\bar{A}_t,\bar{B}_t)\), and the explicit \((C_t,D)\)-based readout. The
resulting block remains recurrent and input-dependent, with computation
organized around projections and element-wise operations. This reduced
operator set is better suited to shared MAC reuse, fixed-point quantization,
and fine-grained pipelining: the projection layers share a matrix-vector
datapath, while state update and output gating stream through regular
arithmetic and activation stages, as described in
Section~\ref{sec:reconfigurable_hw}.

The temporal window length $K$ was evaluated to characterize its impact on localization accuracy and hardware buffering requirements. As shown in Table~\ref{tab:k_ablation_tcas}, increasing the temporal context does not improve the test accuracy, with the lowest test errors obtained for $K=1$ and $K=2$. These results indicate that short temporal windows are sufficient for the considered localization task, while also reducing the buffering requirements of the hardware implementation.
\begin{table}[!t]
\caption{Representative temporal-window ablation for Slim-Mamba.}
\label{tab:k_ablation_tcas}
\begin{center}
\footnotesize
\setlength{\tabcolsep}{6pt}
\renewcommand{\arraystretch}{1.08}
\begin{tabular}{|c|c|c|}
\hline
\textbf{\(K\) (\(P\), \(S\))}
& \textbf{Validation ME (cm)}
& \textbf{Test ME (cm)} \\
\hline
1 (1,1) & 10.07 & \textbf{13.50} \\
\hline
2 (2,2) & 9.56 & \textbf{13.50} \\
\hline
8 (2,2) & 9.41 & 14.10 \\
\hline
4 (4,4) & 9.98 & 13.70 \\
\hline
8 (4,4) & 9.75 & 14.40 \\
\hline
8 (8,4) & 9.74 & 15.20 \\
\hline
32 (8,4) & 9.33 & 16.60 \\
\hline
\end{tabular}
\end{center}
\end{table}

Table~\ref{tab:model_variants_tcas} compares the localization accuracy, model
size, and training cost of the evaluated models. The FCNN provides a published
accuracy reference, whereas the Mamba-family models are evaluated on LuViRA
to select a compact deployment target. Unreported values are marked as N/A.

\begin{table*}[!t]
\begin{center}
\caption{Software model comparison.}
\label{tab:model_variants_tcas}
\footnotesize
\setlength{\tabcolsep}{3pt}
\renewcommand{\arraystretch}{1.08}
\begin{tabular}{|p{0.18\textwidth}|p{0.18\textwidth}|>{\centering\arraybackslash}p{0.12\textwidth}|>{\centering\arraybackslash}p{0.15\textwidth}|>{\centering\arraybackslash}p{0.12\textwidth}|>{\centering\arraybackslash}p{0.12\textwidth}|}
\hline
\textbf{Model} & \textbf{Input features} & \textbf{Model file size (MB)} & \textbf{Avg. training time / epoch (s)} & \textbf{Train rate (batch/s)} & \textbf{Mean localization error (cm)} \\
\hline
FCNN~\cite{LuViRAValidation2024} & CIR & 789 & N/A & N/A & \(\sim 13\) \\
\hline
CMamba~\cite{zeng2024cmamba} & CIR + Power & 40.4 & 361 & 10.31 & 17.4 \\
\hline
Original Mamba & CIR + Power & 3.38 & 33 & 32.62 & 24.74 \\
\hline
Slim-Mamba & CIR + Power & 3.40 & 13 & 75.30 & 13.5 \\
\hline
\end{tabular}
\end{center}
\end{table*}

The FCNN entry is the published reference result from~\cite{LuViRAValidation2024}
and is included only as an accuracy and model-size reference. The Mamba-family
models were trained in PyTorch on an Intel Core i5-13600KF host with an
NVIDIA RTX~3080 GPU. The comparison uses the best evaluated configuration of
each model, with model-specific training--validation splits, temporal windows,
and sampling settings.
In Table~\ref{tab:model_variants_tcas},
``Avg. training time / epoch'' denotes the measured elapsed time required to
complete one training epoch on the same hardware platform, excluding validation
and checkpoint saving,
``Train rate'' denotes the corresponding mini-batch processing rate,
and ME is defined in Section~\ref{sec:system_model}. Model file sizes include
serialization metadata and FP32 parameters for the Mamba-family models.
Table~\ref{tab:slim_param_breakdown_tcas} reports the deployed parameter count.

Slim-Mamba achieves the lowest ME among the evaluated Mamba-family models
and the shortest measured training time per epoch, with a file size comparable
to original Mamba (3.40 versus 3.38\,MB). Its hardware advantage lies in
the smaller recurrent state and simpler operator set, which reduce state
storage and simplify the fixed-point datapath.
Relative to CMamba, it reduces
the model file size from 40.4 to 3.40\,MB and ME from 17.4 to 13.5\,cm.

The odd/even split places test trajectories between nearby training
trajectories. The reported errors therefore characterize spatial
interpolation within the measured grid scenario.

\begin{figure}[!t]
  \centering
  \includegraphics[width=0.96\linewidth]{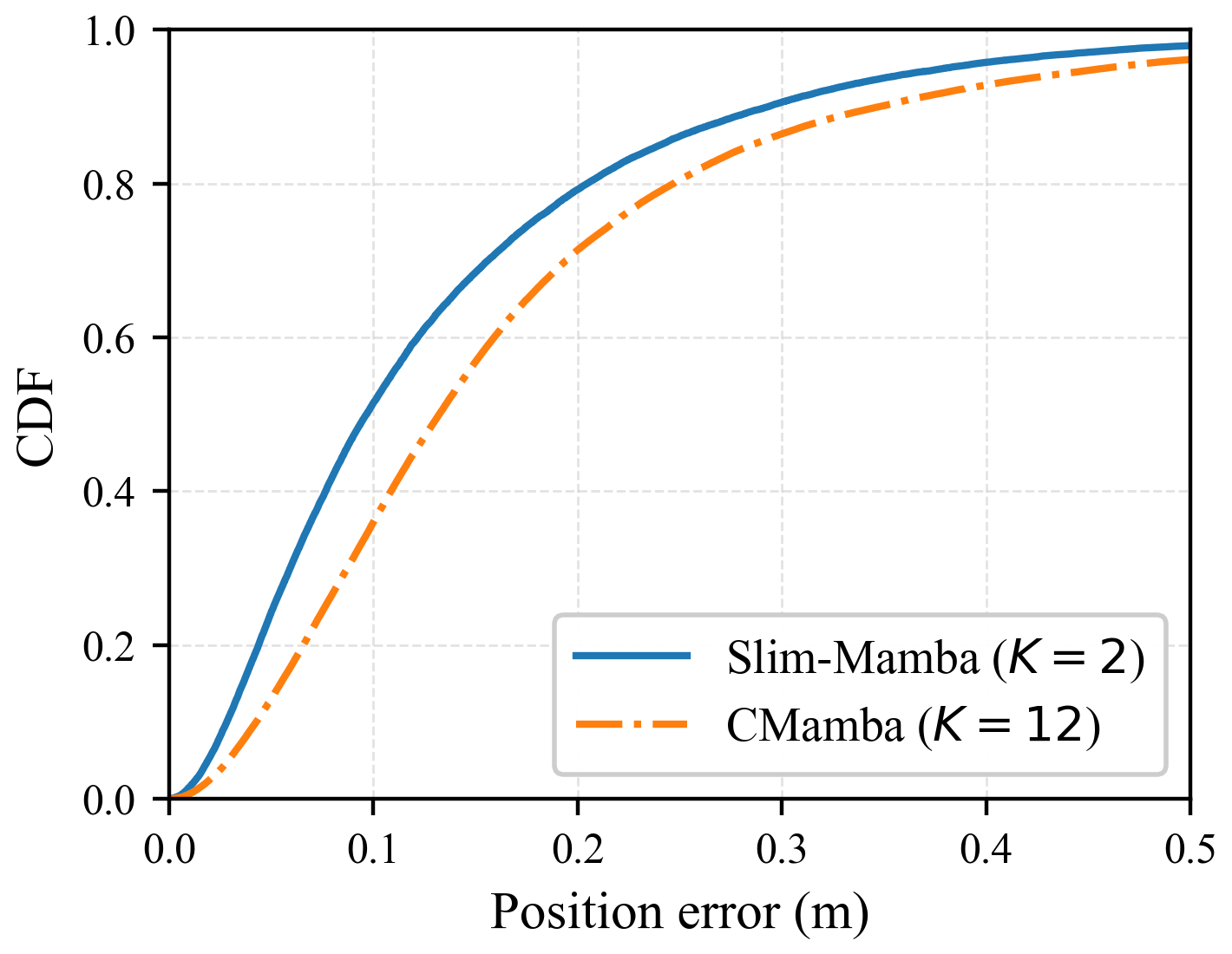}
  \caption{Localization-error CDFs on held-out grid trajectories.}
  \label{fig:model_tradeoff_tcas}
\end{figure}

Figure~\ref{fig:model_tradeoff_tcas} shows the saved error distributions of
Slim-Mamba and CMamba on the same 33 held-out even-indexed grid trajectories,
both using CIR and power features.

\subsection{Deployed Configuration}

The deployed Slim-Mamba configuration is summarized in
Table~\ref{tab:slim_config_tcas}. The selected setting preserves the compact
short-window behavior observed in software while keeping the projection and
state-update dimensions compatible with the shared internal vector widths and
buffering structure of the target FPGA accelerator.

\begin{table}[!t]
\caption{Deployed Slim-Mamba configuration and parameter roles.}
\label{tab:slim_config_tcas}
\begin{center}
\footnotesize
\setlength{\tabcolsep}{4pt}
\renewcommand{\arraystretch}{1.08}
\begin{tabular}{|c|c|p{0.47\linewidth}|}
\hline
\textbf{Symbol} & \textbf{Value} & \textbf{Role in the deployed model} \\
\hline
\(D_{\mathrm{in}}\) & 2100 & Input feature dimension for each radio observation. \\
\hline
\(K\) & 2 & Temporal window length, i.e., the number of consecutive observations processed in one inference window. \\
\hline
\(\text{proj\_dim}\) & 64 & Feature dimension after the input projection and before the final regression head. \\
\hline
\(d_{\mathrm{model}}\) & 128 & Internal hidden dimension used inside each Slim-Mamba block. \\
\hline
\(n_{\mathrm{layer}}\) & 4 & Number of stacked Slim-Mamba blocks in the deployed model. \\
\hline
\(P\) & 2 & Patch length used in the Conv1d patch-embedding stage. \\
\hline
\(S\) & 2 & Patch stride used in the Conv1d patch-embedding stage; \(P=S=2\) yields non-overlapping patches. \\
\hline
\end{tabular}
\end{center}
\end{table}

Table~\ref{tab:slim_param_breakdown_tcas} summarizes the trainable parameters.
Most belong to the input, update-coefficient, and output projections within
the four Slim-Mamba blocks. These projections map to the shared MAC array;
the remaining recurrent update uses element-wise operations.

\begin{table}[!t]
\caption{Trainable parameter breakdown of the Slim-Mamba configuration.}
\label{tab:slim_param_breakdown_tcas}
\begin{center}
\footnotesize
\setlength{\tabcolsep}{6pt}
\renewcommand{\arraystretch}{1.08}
\begin{tabular}{|l|r|}
\hline
\textbf{Component} & \textbf{Parameters} \\
\hline
Input projection \(D_{\mathrm{in}}\rightarrow\text{proj\_dim}\) & 134,464 \\
\hline
Patch embedding Conv1d & 16,512 \\
\hline
Four block RMSNorm layers & 512 \\
\hline
Four block input-projection layers & 262,144 \\
\hline
Four block update-coefficient projection layers & 263,168 \\
\hline
Four block output-projection layers & 131,072 \\
\hline
Final RMSNorm & 128 \\
\hline
Flat output layer & 8,256 \\
\hline
Regression head \(\text{proj\_dim}\rightarrow 2\) & 130 \\
\hline
\textbf{Total} & \textbf{816,386} \\
\hline
\end{tabular}
\end{center}
\end{table}

\section{Reconfigurable Slim-Mamba Accelerator Architecture}
\label{sec:reconfigurable_hw}

This section presents the hardware architecture of the proposed Slim-Mamba accelerator. The proposed architecture supports reconfigurability, improving hardware utilization and enabling flexible adaptation to different computational and numerical requirements. 

\begin{figure*}[!t]
\centering
\includegraphics[width=0.95\textwidth]{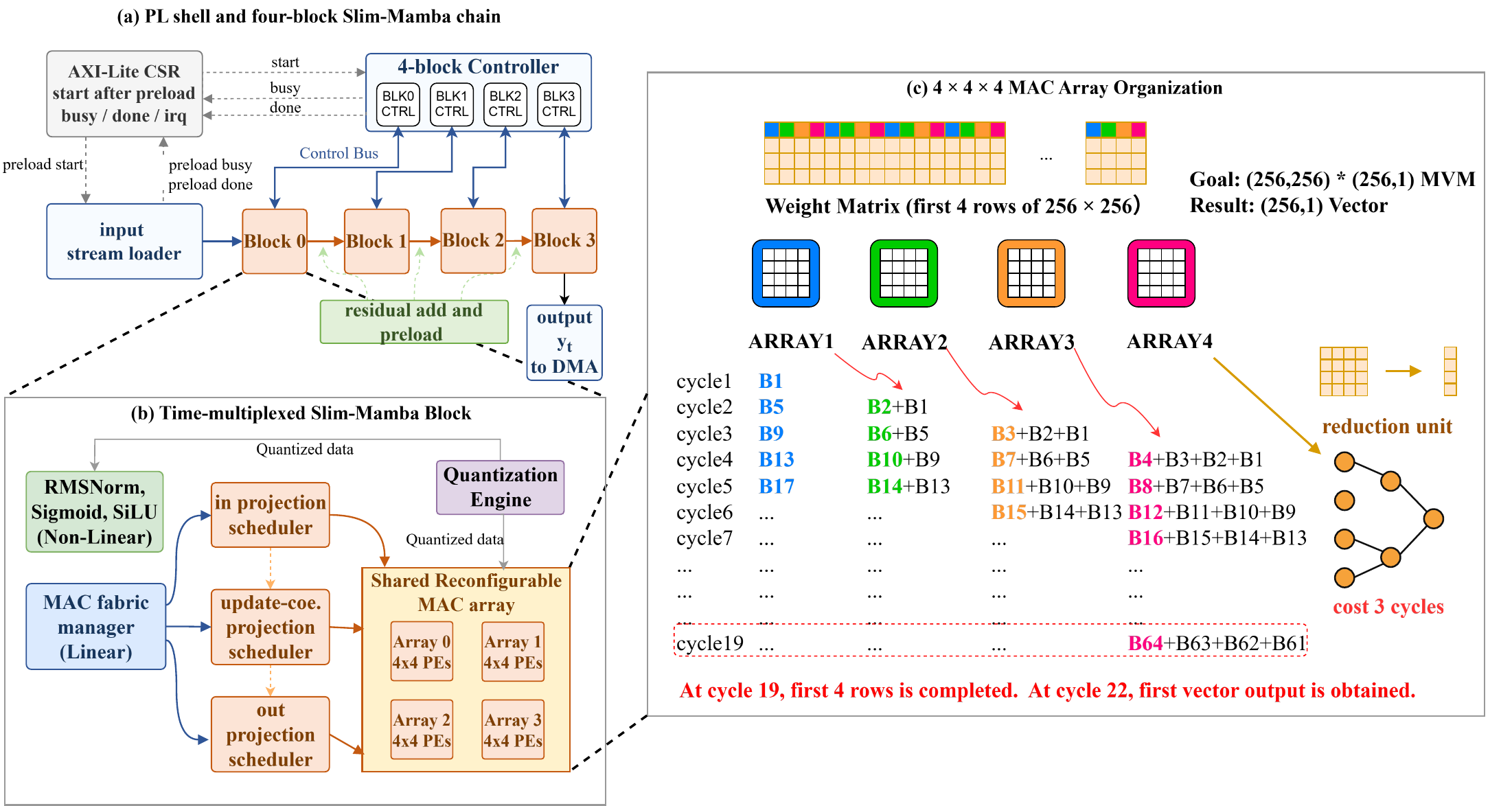}
\caption{Reconfigurable Slim-Mamba accelerator architecture: (a) PL shell and four-block Slim-Mamba
chain, (b) time-multiplexed Mamba block with shared projection fabric, and (c) $4\times4\times4$ MAC Array Organization.}
\label{fig:hardware_architecture}
\end{figure*}

Fig.~\ref{fig:hardware_architecture} shows the overall architecture. The PL-side shell connects the AXI interfaces to the four-block Slim-Mamba
compute chain, as shown in Fig.~\ref{fig:hardware_architecture}(a). The processor-side control logic initializes the accelerator and loads the AXI-Stream data into the buffer of Block~0. Each block implements one full Slim-Mamba block. The four-block chain supports two inter-block transfer modes: a streaming mode that forwards block outputs for inter-block pipelining, and a preload mode allowing each block to execute sequentially, so intermediate results can be inspected and compared before the next block starts.

\subsection{Reconfigurable MAC Array}

Inside each Slim-Mamba block, the main MAC-intensive operators are the input
projection, the update-coefficient projection, and the output
projection. Although these operators use different input buffers, weight banks,
and output destinations, they can all be represented as matrix-vector
multiplication. The accelerator therefore avoids instantiating independent MAC arrays for
these projection layers. Instead, it uses
one shared $4\times4\times4$ MAC array, as illustrated in
Fig.~\ref{fig:hardware_architecture}(b).

Although all projection layers perform matrix-vector multiplication and use the same MAC array to do the calculation, their varying dimensions require dedicated schedulers for data tiling. The
input-projection scheduler reads the block input or hidden representation and
the input-projection weights, and converts them into activation and weight tiles.
The update-coefficient scheduler reads the Slim-Mamba state-update input cache and the update-coefficient
projection weights, and generates the corresponding tile stream. The
output-projection scheduler reads the gated state-update output and the output-projection
weights. Although these three schedulers access different buffers and write to
different destinations, they expose the same fabric interface. A MAC array
manager arbitrates among the schedulers and configures the shared fabric to
serve one projection role at a time.

The projection stages within each Slim-Mamba block share a single MAC array, reducing DSP duplication and placement overhead at the cost of limited projection-level concurrency. The fabric interface includes control parameters that determine the computation dimensions, allowing the $4\times4\times4$ array to support
different logical matrix shapes, reduction lengths, and output forms. 


The $4\times4\times4$ fabric contains four $4\times4$ sub-arrays. Each
sub-array receives a four-element activation slice and a $4\times4$ weight
block. The four sub-arrays operate with a staggered schedule and their partial
results are combined through a reduction tree. Fig.~\ref{fig:hardware_architecture}(c)
illustrates the array organization for the update-coefficient projection in the
state-update path. The symbol B in the table denotes the current $4\times4$
input tile. The weight matrix shown below represents a portion of the
update-coefficient weight matrix, where each square corresponds to one
$4\times4$ block. The four sub-arrays operate in a staggered pipeline. At cycle
19, the first four rows of the weight matrix have been fully consumed, and at
cycle 22 the first output tile is produced after the reduction tree.
 
The PE-array organization should be matched to the data-supply and reduction pattern of the 
Slim-Mamba workload. Systolic arrays are a classical and widely used organization for regular 
matrix computations ~\cite{Kung1982Systolic,Liu2020STA}, while previous studies 
have shown that data movement, local reuse, and dataflow mapping are critical to energy efficiency 
and throughput~\cite{Sze2017EfficientDNN,Chen2017Eyeriss}. However, systolic arrays are fundamentally optimized for dense general matrix-matrix multiplication (GEMM). Because the main computation in Slim-Mamba projection 
operators is MVM (matrix-vector multiplication), our $4\times4\times4$ arrays organization reduces global routing and
data-supply pressure for the MVM-dominated workload compared with a systolic array. Compared with a
$64\times1$ GEMV engine, which is designed for MVM calculations, it consumes the same amount of hardware resources but provides
earlier tile availability. Still take Slim-Mamba state update datapath as example, which
performs MVM with the matrix shape of (256,256) and vector shape of (256,1). With the 
four $4\times4$ sub-arrays working in pipeline, the first output vector can be produced 
after 22 cycles rather than waiting for a full
vector-reduction interval in a $64\times1$ GEMV engine, which is 256 cycles. 
This earlier tile output is important because the
following nonlinear and state-update stages can start before the whole
projection vector is completed.

\subsection{Fine-Grained FIFO-Based Pipeline}

As discussed above, the PEs inside the $4\times4\times4$ MAC array can be configured
for MVM(Matrix Vector Multiplication), EWM(Element-wise Multiplication), and EWA(Element-wise Addition), so the tradeoff between hardware reuse and latency must be considered.
A coarse implementation of the state-update path would execute update-coefficient projection,
sigmoid, state update, SiLU, and output gating sequentially. This maximizes
resource reuse, but it forces each stage to wait until the previous stage has
produced a full vector. The proposed accelerator instead uses a fine-grained
FIFO-based pipeline. Once the MAC array produces and reduces an output tile,
the tile is forwarded through FIFO interfaces to the bias-add, sigmoid, and
state-update stages. The state-update unit starts as soon as the corresponding
tile, update coefficient, state value, and scale are available. It does not wait
for the full projection vector.

The state update follows the Slim-Mamba recurrence
in
Eq.~\eqref{eq:slim_mamba_state_update_alg}, where $\lambda_t$ is generated by the
update-coefficient projection followed by the sigmoid unit. The updated state is then
element-wise multiplied by the SiLU gate output. FIFOs are
used to align the projection output, sigmoid value, gate path, state read result,
and dynamic scale. This alignment is necessary because different sub-stages have
different local latencies.

\begin{figure}[!t]
\centering
\includegraphics[width=0.98\linewidth]{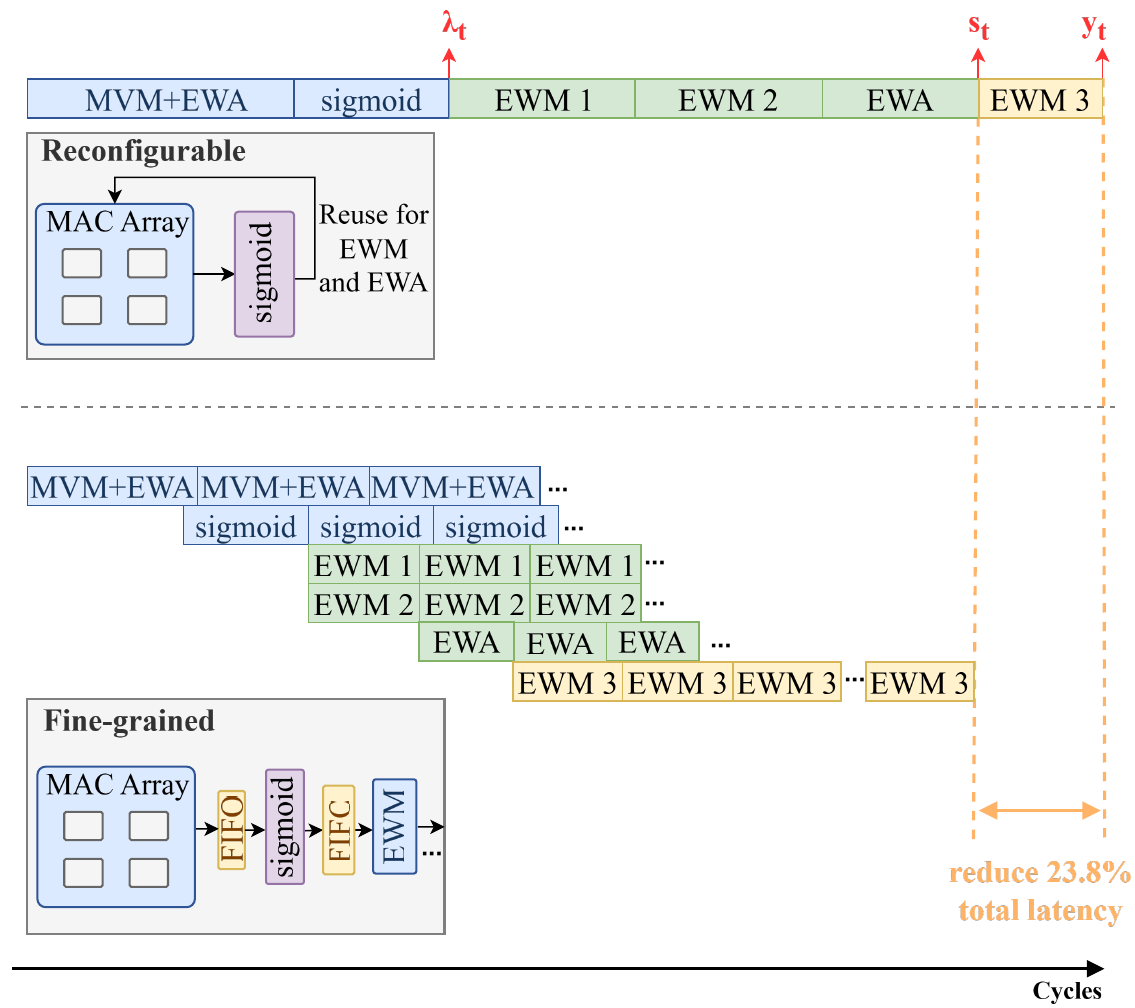}
\caption{Comparison between coarse-grained and fine-grained scheduling.}
\label{fig:fine_grained_pipeline}
\end{figure}

\begin{figure}[!t]
\centering
\includegraphics[width=0.98\linewidth]{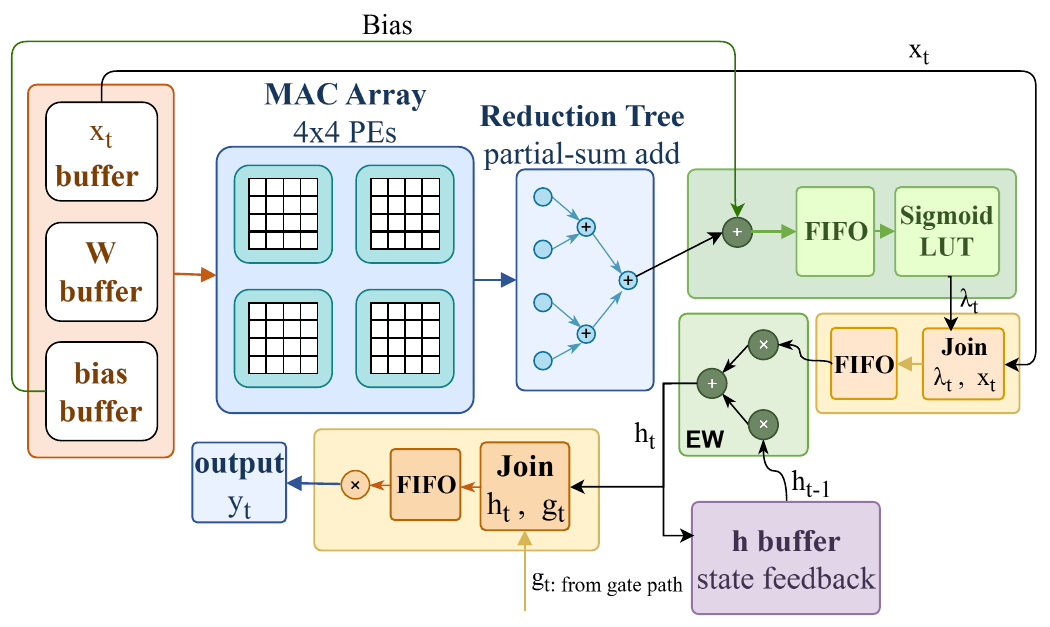}
\caption{Slim-Mamba state-update datapath.}
\label{fig:ssm_path}
\end{figure}

For an $N$-tile Slim-Mamba state-update-path workload, the coarse-grained schedule can be abstracted as
\begin{equation}
T_{\mathrm{coarse}} = 29N ,
\label{eq:coarse_latency}
\end{equation}
where projection, nonlinear activation, state update, and output gating are
mostly serialized. With fine-grained FIFO streaming, the schedule becomes
\begin{equation}
T_{\mathrm{fine}} = 22N + 7 .
\label{eq:fine_latency}
\end{equation}
For $N=64$, this reduces the state-update-path latency from 1856 cycles to 1415 cycles,
corresponding to a 23.8\% reduction. The additional cost is about 12 DSPs per
state-update path, which is acceptable compared with the latency reduction.
Figure~\ref{fig:fine_grained_pipeline} shows the comparison. Fig.~\ref{fig:ssm_path}
shows the final fine-grained Slim-Mamba state-update datapath, where the hardware modules implementing the operations in Eqs.~\eqref{eq:slim_mamba_lambda}, \eqref{eq:slim_mamba_state_update_alg}, \eqref{eq:slim_mamba_gate}, and~\eqref{eq:slim_mamba_output}, including recurrent state update, bias addition, sigmoid, SiLU gating, and output generation, are connected through FIFO-based interfaces.

\subsection{Quantization Strategy and Hardware implementation}

As the quantization strategy depends on the data distribution, we first profile the range of floating-point intermediate values in the trained Slim-Mamba model using 4096 representative input samples. For each layer, following Slim-Mamba algorithm in Fig.~\ref{fig:slim_mamba_replacement}, we analyze the maximum absolute output values. Since symmetric quantization is used to map values symmetrically around zero without a zero-point offset and simplify the hardware datapath, this maximum determines the scale and directly affects the quantization resolution. 

\begin{figure}[!t]
\centering
\includegraphics[width=0.98\linewidth]{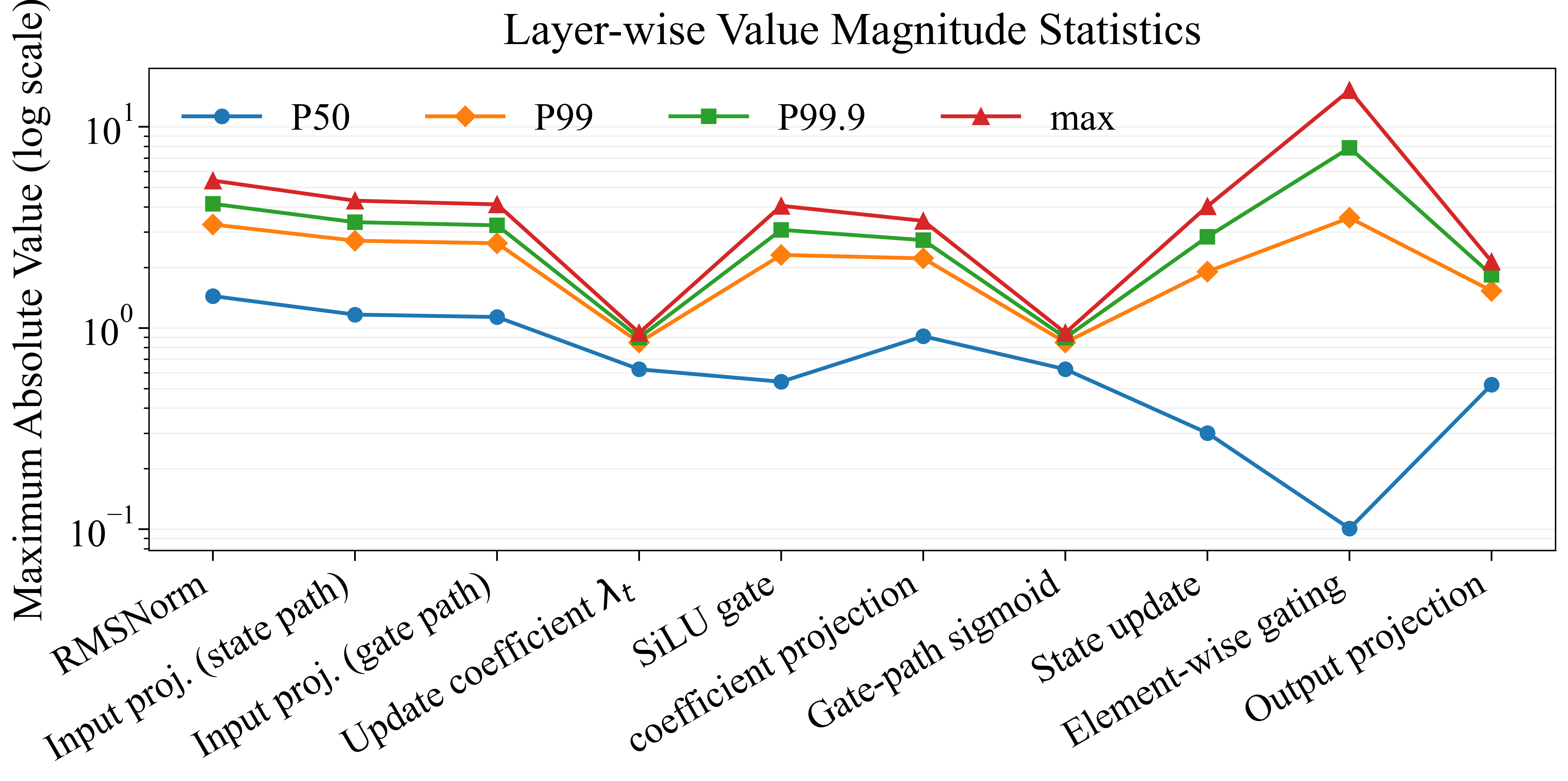}
\caption{Layer-wise Value Magnitude Statistics}
\label{fig:datarange}
\end{figure}

In Fig.~\ref{fig:datarange}, P50, P99, P99.9, and max denote the distribution of the maximum absolute output value. The profiling result shows that the required numerical range is strongly
layer-dependent. The update coefficient and sigmoid layers have a compact range,
while the recurrent state and element-wise gating layers show a much larger gap
between the median and tail values. In particular, the state-update path and corresponding gating path has a
small median range but significantly larger P99 and maximum values, indicating a
long-tailed distribution across channel groups and input samples. Therefore, when determining the scaling strategy, the state-update layer requires
a dedicated quantization configuration to accommodate its dynamic data range. The strategy is called dynamic scaling, which means the scale factor is computed from the current recurrent-state input magnitude and used when storing and restoring the state. This allows the state representation to adapt to data ranges that vary over
time.

Table~\ref{tab:stage_quant_error} shows the experimental results for scaling strategy. We first evaluate a fixed-scale baseline because it avoids runtime dynamic scale
generation, per-layer scale storage, and scale-alignment control. In this baseline, fixed-point data are represented in Q8.8,
where the fixed scale is \(s=2^{-8}=1/256\) and
\(x_{\mathrm{real}}=s\cdot x_{\mathrm{int}}\). Q8.8 is used as the fixed-scale baseline because it 
offers a balanced integer/fractional split.
The fixed-scale baseline gives MAEs of 0.00--0.01 before the recurrent
state update. In contrast, the state-update MAE reaches
0.13, which validates the profiling result in Fig.~\ref{fig:datarange}. After dynamic scaling is applied to the state update path, the
MAE is reduced to 0.09.

\begin{table}[!t]
\begin{center}
\caption{Stage-Level Quantization Error With Dynamic Scaling Applied to the State Path.}
\label{tab:stage_quant_error}
\footnotesize
\setlength{\tabcolsep}{6pt}
\begin{tabular}{|l|c|c|}
\hline
\textbf{Stage} & \textbf{Fixed-scale MAE} & \textbf{Dynamic-scale MAE} \\
\hline
RMSNorm & 0.00 & 0.00 \\
\hline
Input proj. (state path) & 0.01 & 0.01 \\
\hline
Input proj. (gate path) & 0.01 & 0.01 \\
\hline
Update coefficient $\lambda_t$ & 0.01 & 0.01 \\
 \hline
 Output-gate SiLU & 0.01 & 0.01 \\
 \hline
 coefficient projection & 0.01 & 0.01 \\
 \hline
 Update sigmoid ($\lambda_t$) & 0.00 & 0.00 \\
 \hline
 \textbf{State update} & \textbf{0.13} & \textbf{0.09} \\
 \hline
 Element-wise gating & 0.03 & 0.03 \\
 \hline
 Output projection & 0.03 & 0.03 \\
 \hline
 \end{tabular}
 \end{center}
 \end{table}

After identifying the scaling strategy, the exact Q format, rounding mode,
and saturation mode for each layer are selected through statistics-guided simulation search. The
software search first inserts quantize--dequantize operations at layer
boundaries to estimate the numerical loss of each candidate configuration~\cite{Jacob2018IntegerOnly,Krishnamoorthi2018Quantization}. The
promising candidates are then verified using a bit-true C++ model that follows
the RTL behavior, including binary-point shifts, round-to-nearest-even,
saturation, and runtime state-scale
conversion. 

\begin{figure*}[!t]
\centering
\includegraphics[width=0.90\textwidth]{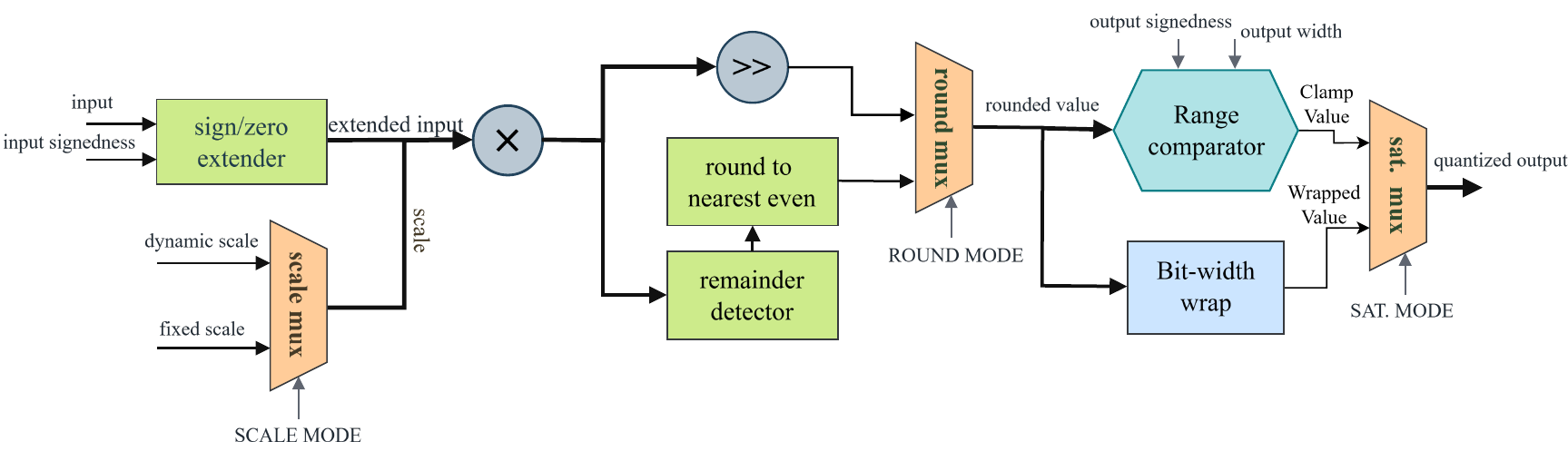}
\caption{Configurable fixed-point requantization engine reused across
projection, state-update, and residual paths.}
\label{fig:requant_engine}
\end{figure*}

The quantization strategy is implemented by a shared reconfigurable requantization
engine rather than by instantiating a separate quantization unit for every
layer. This module is used
after MAC accumulation, Q-format conversion, state update, scale multiplication,
and residual addition. As shown in Fig.~\ref{fig:requant_engine}, this shared structure reduces duplicated hardware and allows the layer-wise quantization policy obtained from software
analysis to be adjusted easily to hardware control parameters.

\begin{figure}[!t]
\centering
\includegraphics[width=0.95\linewidth]{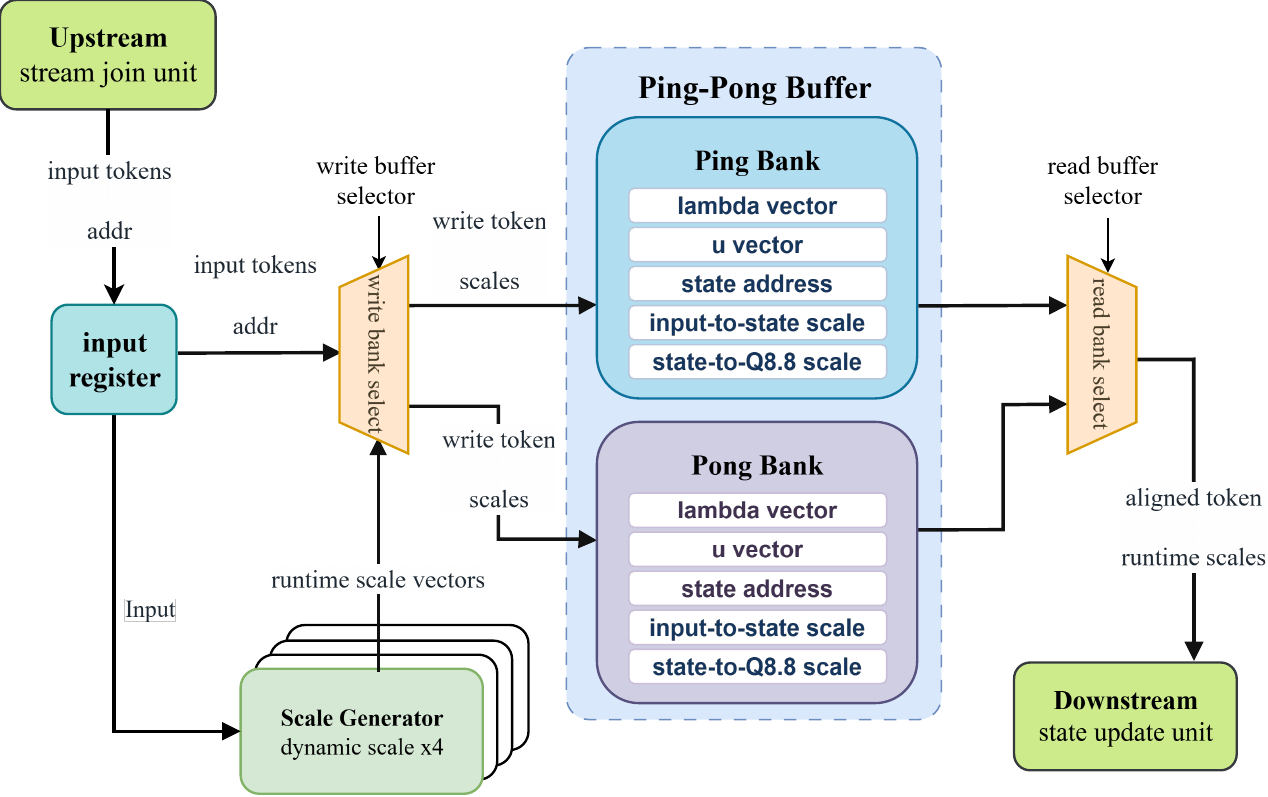}
\caption{Ping-pong scale-buffer organization for aligning token, state address,
update coefficient, and runtime scale streams.}
\label{fig:ping_pong}
\end{figure}

The dynamic-scale state-update path requires additional alignment logic. Since the
state-update datapath is stream based, each token, state address, update
coefficient, and runtime scale must arrive at the state-update unit in the same
order. A ping-pong scale buffer is used for this purpose, as seen in
Fig.~\ref{fig:ping_pong}.
While one bank receives the incoming tokens, their state addresses, and the
generated runtime scales, the other bank provides an already aligned
token-scale pair to the downstream state-update pipeline. The write-buffer
selector and read-buffer selector identify the active bank. The read and write
banks switch between consecutive tokens, which hides the
scale-generation latency and preserves streaming order. In this way, dynamic
scaling can be inserted into the state-update path without breaking the
fine-grained FIFO-based pipeline.

The final fixed-point model achieves an
output MAE of 0.02 with respect to the floating-point output and a mean
localization error of 0.14~m, which is close to the
floating-point reference,
showing that the selected layer-wise fixed-point strategy preserves the
localization accuracy while remaining compatible with the hardware datapath.


\subsection{Nonlinear Layer Implementation}

From the algorithm analysis in Section~\ref{sec:algorithm}, the main nonlinear operators in
the Slim-Mamba model are RMSNorm, sigmoid, and SiLU.

RMSNorm is placed before the projection path to stabilize the dynamic range of
the input features. For an input vector
$\mathbf{x}=[x_0,x_1,\ldots,x_{D-1}]$, the RMS value is
\begin{equation}
\mathrm{RMS}(\mathbf{x})=
\sqrt{\frac{1}{D}\sum_{i=0}^{D-1}x_i^2+\epsilon},
\label{eq:rms_value}
\end{equation}
and the normalized output is computed as
\begin{equation}
\hat{x}_i =
\gamma_i x_i \cdot
\widehat{\frac{1}{\mathrm{RMS}(\mathbf{x})}} .
\label{eq:rmsnorm_output}
\end{equation}
where $\gamma_i$ is the learned channel-wise gain and the reciprocal is
approximated in fixed point.
The square-root operation is implemented with a trial-quotient digit-recurrence
procedure, which uses shift, comparison, and subtraction operations instead of a
general floating-point square-root unit. This design follows 
the general principle of digit-recurrence square-root methods, where the root is formed iteratively 
through residual comparison and root-digit selection ~\cite{Harris2024Radix4Sqrt,Bruguera2023Radix64Sqrt}. The final 
division is replaced by reciprocal generation followed by multiplication. A Q30 reciprocal 
is used so that the shared normalization scale remains accurate for all channels of one
token.

The sigmoid function is used to generate the state-update coefficient: 
\begin{equation} 
  \sigma(x)=\frac{1}{1+e^{-x}}. \label{eq:sigmoid} 
\end{equation} 
Direct evaluation of the exponential function is costly in fixed-point digital hardware. 
Therefore, lookup-table-based approximations have been widely used for hardware implementations of 
sigmoid and other neural-network activation functions~\cite{Meher2010SigmoidLUT,Piazza1993DigitalLUT}. 
In this work, the sigmoid input is clamped to $[-4,+4)$, and a 2048-entry LUT stores the corresponding Q0.16 sigmoid values. 
This avoids evaluating the exponential function in hardware. SiLU reuses the same sigmoid path:
\begin{equation}
\mathrm{SiLU}(x)=x\cdot\sigma(x).
\label{eq:silu}
\end{equation}
The original input is delayed through a FIFO while the sigmoid value is produced.
After alignment, an element-wise multiplier generates the SiLU output. Therefore,
sigmoid and SiLU share the clamp and LUT logic, while SiLU only adds FIFO
alignment and vector multiplication.

\section{Results}
This section reports fixed-point accuracy, core throughput, FPGA resource
utilization, and power, followed by the effect of fine-grained pipelining.
Table~\ref{tab:accelerator_compare_tcas} compares representative localization
implementations and their measurement scenarios.

\begin{table*}[!t]
\caption{Contextual comparison of massive MIMO-based localization models and corresponding accelerators.}
\label{tab:accelerator_compare_tcas}
\begin{center}
\footnotesize
\begin{tabular}{|c|c|c|c|c|c|c|c|}
\hline
\textbf{Work} & \textbf{Model} & \textbf{Dataset / scenario} &
\textbf{Platform} & \textbf{Clock} & \textbf{Arith.} &
\textbf{Position/s} & \textbf{ME (m)} \\
\hline
\cite{LuViRAValidation2024}
& FCNN & LuViRA, Indoor & CPU & N/A & Float & N/A
& \(\sim 0.13\) \\ \hline

\cite{AttariISCAS2021}
& CNN & COST 2100, & 22 nm ASIP + & 555\,MHz
& INT16 & 271 & 0.4$^{\mathrm{a}}$ \\
&& simulated & CNN accelerator &&&& \\ \hline

\cite{YamanTransformer2026}
& Transformer & Commercial, Outdoor & FPGA & 100\,MHz
& INT16 & up to 1961 & $<1$ \\ \hline

\cite{CastanedaSIXPPAC2026}
& SIXPPAC & DICHASUS, & 12 nm ASIC & 840\,MHz
& W1A8$^{\mathrm{c}}$ & $105\times10^{6}$ & 0.46, 0.69, \\
& & Indoor, Outdoor &&&&& 2.13$^{\mathrm{d}}$ \\ \hline

\textbf{This work}
& Slim-Mamba & LuViRA, Indoor & FPGA & 100\,MHz
& INT16 & 4,716$^{\mathrm{b}}$ & 0.14 \\ \hline

\multicolumn{8}{l}{$^{\mathrm{a}}$Reported as an average positioning error of $3.5\lambda$ (or root-mean-square error: 40 cm).} \\
\multicolumn{8}{l}{$^{\mathrm{b}}$Estimated four-block core throughput for offline-preprocessed inputs.} \\
\multicolumn{8}{l}{$^{\mathrm{c}}$Binary weights and 8-bit inputs/activations; positioning uses CSI block processing.} \\
\multicolumn{8}{l}{$^{\mathrm{c}}$W1A8 denotes 1-bit binary weights and 8-bit inputs/activations; positioning uses CSI block processing.} \\
\multicolumn{8}{l}{$^{\mathrm{d}}$Mean distance errors on DICHASUS, indoor, and outdoor datasets, respectively; datasets differ across applications.}
\end{tabular}
\end{center}
\end{table*}

\subsection{Throughput, Resource Utilization, and Power}

Using the floating-point model as the reference, the hardware fixed-point implementation reaches an output-coordinate MAE of 0.02\,m and a mean localization error of 0.14\,m. Using the per-block
cycle count in Table~\ref{tab:block_latency_tcas}, serial execution of four
blocks requires $4\times5301=21{,}204$ cycles per token, or
212.04\,\(\mu\)s at 100\,MHz. With $K=P=S=2$, each position window produces
one token, giving an estimated four-block core throughput of
4716\,positions/s for offline-preprocessed inputs.

Table~\ref{tab:resource_utilization_tcas} reports the resource usage of the
shared-MAC implementation within the target FPGA budget.

\begin{table}[!t]
\caption{Post-synthesis resource utilization at 100\,MHz.}
\label{tab:resource_utilization_tcas}
\begin{center}
\footnotesize
\setlength{\tabcolsep}{6pt}
\renewcommand{\arraystretch}{1.08}
\begin{tabular}{|l|c|c|c|}
\hline
\textbf{Resource} & \textbf{Used} & \textbf{Available} & \textbf{Utilization} \\
\hline
LUT & 60,868 & 274,080 & 22.21\% \\
\hline
FF & 77,345 & 548,160 & 14.11\% \\
\hline
BRAM tiles & 279 & 912 & 30.59\% \\
\hline
DSP & 496 & 2,520 & 19.68\% \\
\hline
\end{tabular}
\end{center}
\end{table}

Table~\ref{tab:power_summary_tcas} shows that the processing system (PS)
accounts for 2.62\,W, or 74\% of the estimated dynamic power, compared with
0.94\,W for the programmable logic (PL).

\begin{table}[!t]
\caption{On-chip power estimate.}
\label{tab:power_summary_tcas}
\begin{center}
\footnotesize
\setlength{\tabcolsep}{6pt}
\renewcommand{\arraystretch}{1.08}
\begin{tabular}{|l|c|c|}
\hline
\textbf{Metric} & \textbf{Value} & \textbf{Share} \\
\hline
Total on-chip power & 4.29\,W & 100\% \\
\hline
Dynamic power & 3.56\,W & 83\% \\
\hline
Static power & 0.73\,W & 17\% \\
\hline
PS dynamic power & 2.62\,W & 74\% of dynamic power \\
\hline
PL dynamic power & 0.94\,W & 26\% of dynamic power \\
\hline
\end{tabular}
\end{center}
\end{table}

\subsection{Latency Reduction from Fine-Grained Pipelining}

For the state-update pipeline in Section~\ref{sec:reconfigurable_hw}, the
coarse schedule can be approximated by
\(T_{\mathrm{coarse}}=29N\), whereas the fine-grained schedule becomes
\(T_{\mathrm{fine}}=22N+7\). For \(N=64\) tiles, this reduces the Slim-Mamba state-update path
latency from 1856 cycles to 1415 cycles, corresponding to a 23.8\% reduction.

The per-block latency breakdown is reported in
Table~\ref{tab:block_latency_tcas}. Processing one 128-dimensional token
through one Slim-Mamba block requires 5301 cycles,
or 53.01\(\,\mu\)s at 100\,MHz. Input projection is the dominant stage.
Fine-grained pipelining starts recurrence and nonlinear operations on partial
tiles before the full projection vector is available.

\begin{table}[!t]
\caption{Latency breakdown per token in one Slim-Mamba block.}
\label{tab:block_latency_tcas}
\begin{center}
\footnotesize
\setlength{\tabcolsep}{6pt}
\renewcommand{\arraystretch}{1.08}
\begin{tabular}{|l|c|c|}
\hline
\textbf{Stage} & \textbf{Cycles} & \textbf{Time at 100\,MHz} \\
\hline
Input projection & 2,292 & 22.92\,\(\mu\mathrm{s}\) \\
\hline
State-update path & 1,415 & 14.15\,\(\mu\mathrm{s}\) \\
\hline
Output projection & 1,093 & 10.93\,\(\mu\mathrm{s}\) \\
\hline
Control / alignment overhead & 501 & 5.01\,\(\mu\mathrm{s}\) \\
\hline
\textbf{Total per block} & \textbf{5,301} & \textbf{53.01\,\(\mu\mathrm{s}\)} \\
\hline
\end{tabular}
\end{center}
\end{table}

\section{Conclusions}

This paper presented Slim-Mamba, a hardware-oriented localization model and
reconfigurable FPGA accelerator for massive MIMO indoor localization.
Task-specific model exploration led to a channel-wise gated recurrence
that retains input-dependent state updates and output gating while removing
state expansion, explicit discretization, and input-dependent state readout.
Its recurrence follows the scalar-state special case of the selective SSM,
providing a basis for simplifying the recurrent datapath.

The accelerator maps the resulting projection and element-wise operators
onto a shared MAC datapath and a fine-grained FIFO-based pipeline, with
dynamic scaling for fixed-point recurrent-state storage. On LuViRA, the
hardware-aware fixed-point model achieves a mean localization error of
0.14\,m. Fine-grained pipelining reduces the state-update-path latency by
23.8\%. At 100\,MHz, the four-block core has an estimated serial latency of
212.04\,\(\mu\)s per position window, corresponding to a core throughput of
approximately 4716\,positions/s for offline-preprocessed inputs.

These results demonstrate that task-specific SSM simplification, coupled
with a matched fixed-point architecture, enables low-latency FPGA processing
while preserving decimeter-level localization accuracy.

\bibliographystyle{IEEEtran}
\bibliography{main}

\section*{} 

\vskip -4\baselineskip plus -1fil
\begin{IEEEbiography}
[{\includegraphics[width=1in,height=1.25in,clip,keepaspectratio]{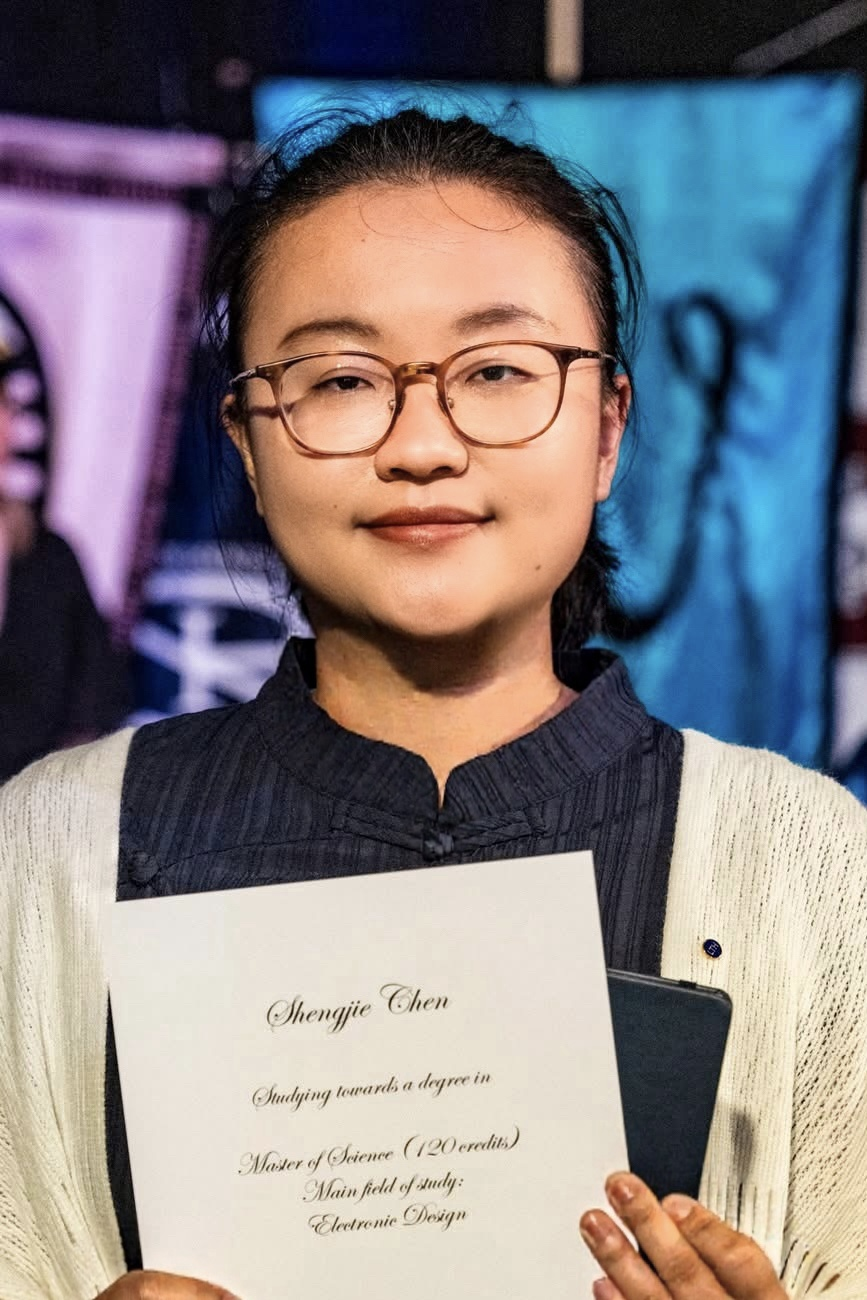}}]
{Shengjie Chen} received the bachelor's degree in Electronic Science and Technology from Southeast University in 2024 and the master's degree in Embedded Electronics Engineering from Lund University in 2026. Her research interests include machine-learning hardware accelerators and digital integrated circuit design.
\end{IEEEbiography}

\vskip -1\baselineskip plus -1fil
\begin{IEEEbiography}[{\includegraphics[width=1in,height=1.25in,clip,keepaspectratio]{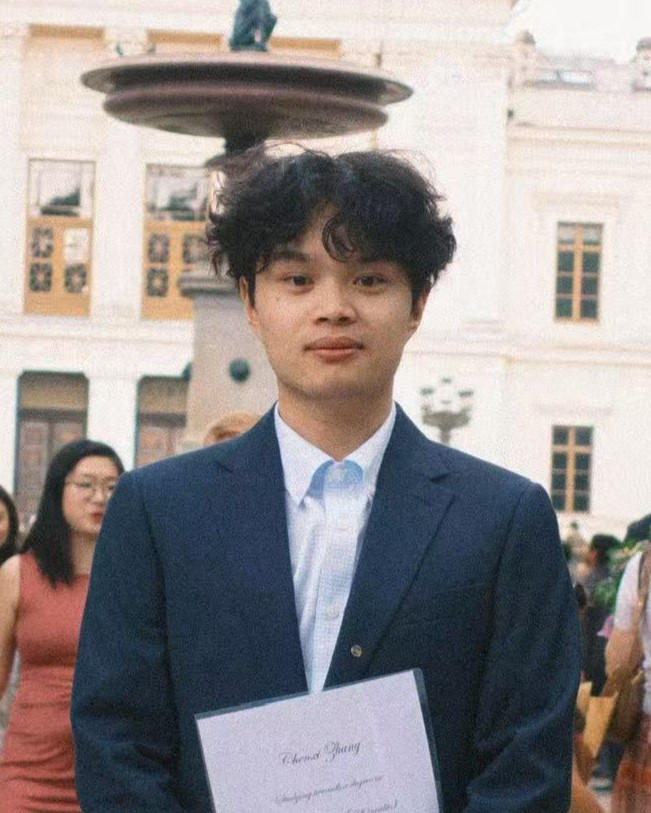}}]
{Chenxi Zhang}
received the bachelor's degree from Southern University of Science and Technology and the master's degree in Embedded Electronics Engineering from Lund University. His research interests include machine-learning hardware deployment and acceleration, especially efficient neural processing unit (NPU) architectures and AI inference accelerators.
\end{IEEEbiography}

\vskip -1\baselineskip plus -1fil
\begin{IEEEbiography}[{\includegraphics[width=1in,height=1.25in,clip,keepaspectratio]{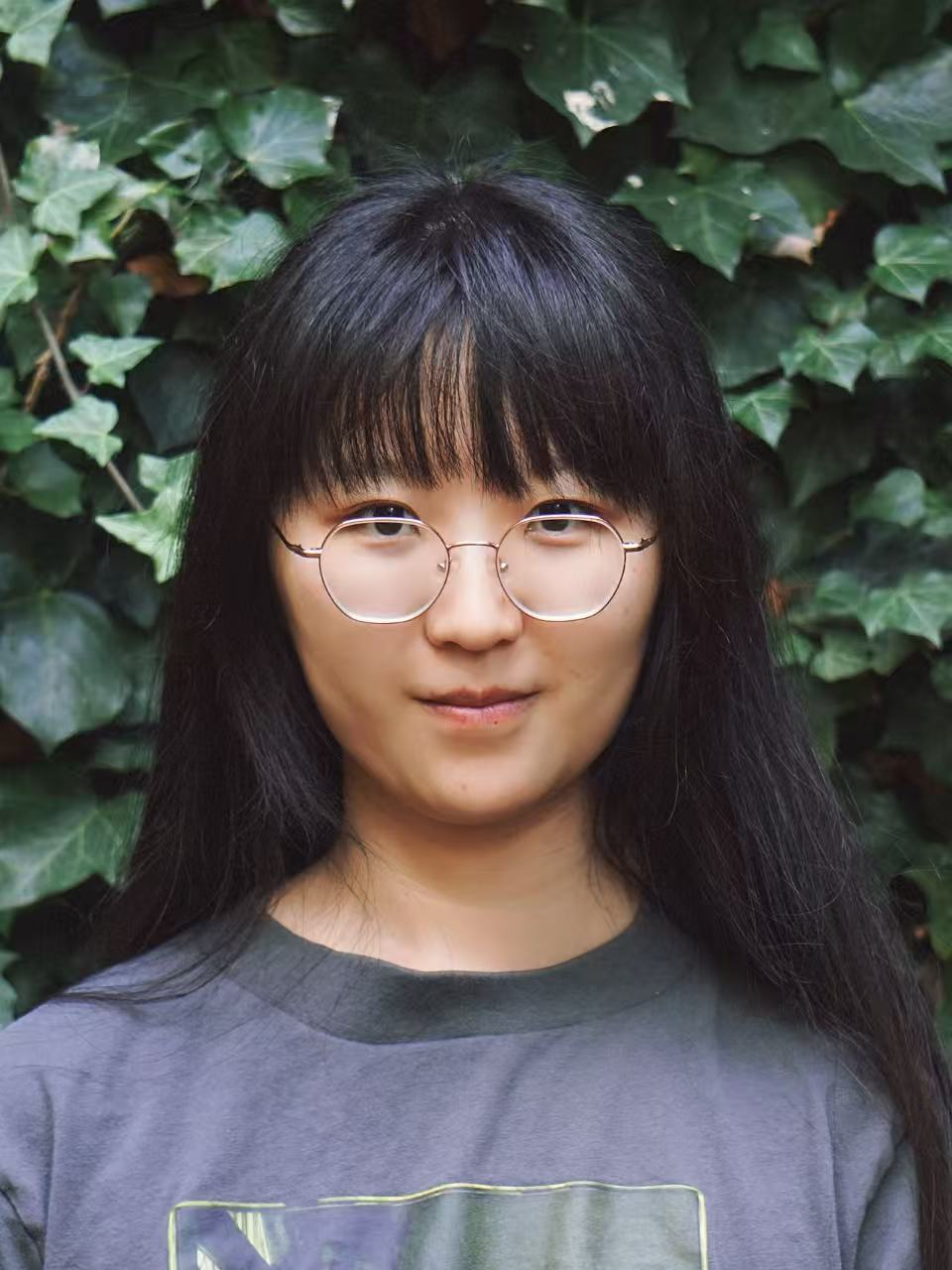}}]{Sijia Cheng}~(Student Member, IEEE) 
received a double bachelor's degree from Beijing Jiaotong University and KU Leuven in 2020, followed by the master's degree in Embedded Electronics Engineering from Lund University in 2022. During her studies at Lund, she received the LU Global Scholarship. She is currently working toward the Ph.D. degree at Lund University under the supervision of Liang Liu. Her research focuses on low-latency digital signal processing for massive MIMO systems.\end{IEEEbiography}

\vskip -1\baselineskip plus -1fil
\begin{IEEEbiography}[{\includegraphics[width=1in,height=1.25in,clip,keepaspectratio]{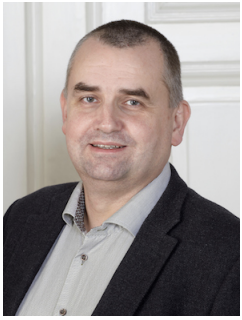}}] {Ove Edfors}~(Senior Member, IEEE) is a Professor of Radio Systems with the Department of Electrical and Information Technology, Lund University, Sweden. His research interests include statistical signal processing and low-complexity algorithms for wireless communications. In the context of massive MIMO and large intelligent surfaces, his current research focuses on how realistic propagation characteristics influence system performance and baseband-processing complexity.
 \end{IEEEbiography}

\vskip -1\baselineskip plus -1fil
\begin{IEEEbiography}[{\includegraphics[width=1in,height=1.25in,clip,keepaspectratio]{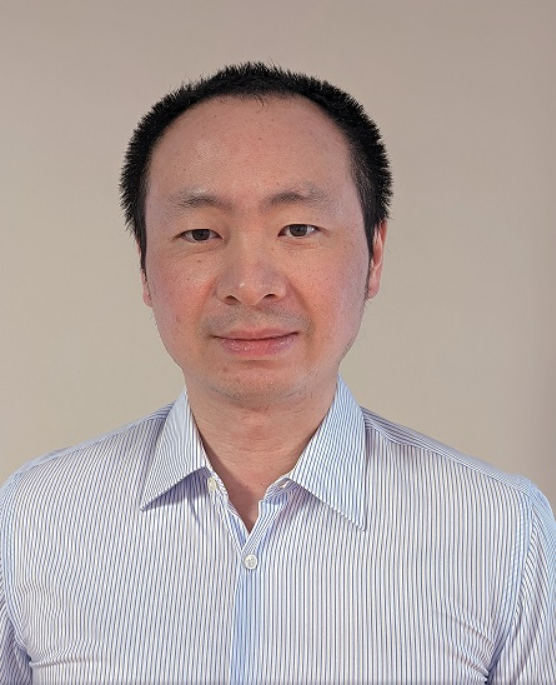}}] {Liang Liu}~(Member, IEEE) received the Ph.D. degree from Fudan University in 2010. He joined Lund University as a postdoctoral researcher. Since 2024, he has been a Professor at Lund University. His research interests include wireless systems and digital integrated circuit design. He is a member of the Technical Committee on VLSI Systems and Applications and the CAS Committee for Communications of the IEEE Circuits and Systems Society.
 \end{IEEEbiography}

\vskip -1\baselineskip plus -1fil
\begin{IEEEbiography}[{\includegraphics[width=1in,height=1.25in,clip,keepaspectratio]{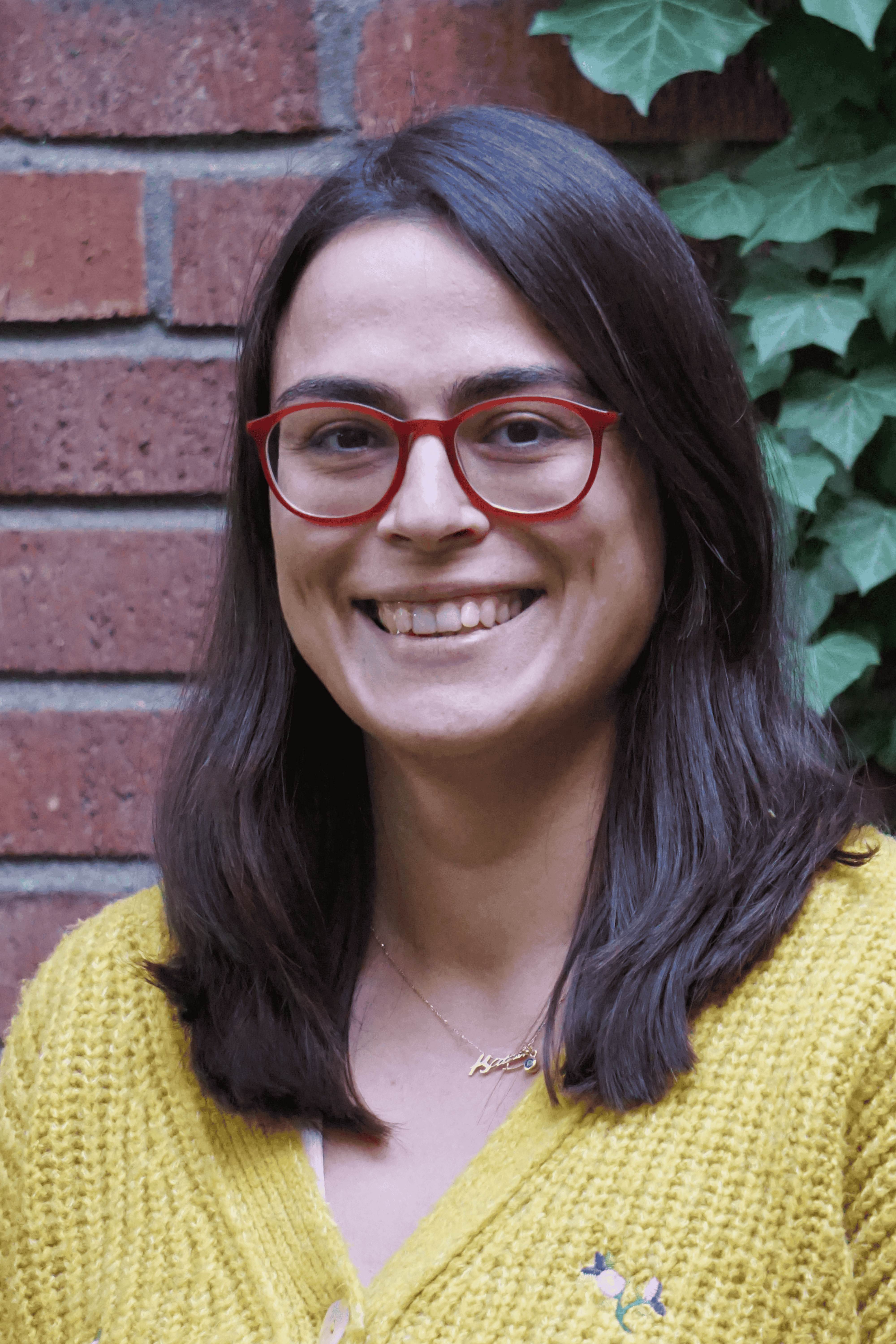}}]
 {Ilayda Yaman}~(Member, IEEE) 
received her bachelor's degree from Istanbul Technical University in 2018 and her master's degree in Embedded Electronics Engineering from Lund University in 2020. During her master's studies, she was awarded the LU Global Scholarship. She completed her Ph.D. degree at Lund University in 2026, under the supervision of Liang Liu. Her current research interests include the efficient implementation of machine learning models in hardware, with a focus on low-power architectures for radio-based localization systems.
\end{IEEEbiography}

\end{document}